\documentclass[prd,superscriptaddress,amsfonts,amssymb,amsmath,showpacs,twocolumn,nofootinbib]{revtex4-2}
\usepackage{bm}
\usepackage{amsfonts}
\usepackage{latexsym}
\usepackage{graphicx}
\usepackage{amsmath}
\usepackage{palatino}
\usepackage{xcolor} 
\usepackage{mathpazo}
\usepackage{tensor}
\usepackage{textcomp}
\usepackage{float}
\usepackage{booktabs}
\usepackage{dcolumn}
\usepackage{booktabs}
\usepackage{multirow}
\usepackage{hyperref}
\hypersetup{colorlinks,citecolor=blue}
\usepackage{amsmath}
\usepackage{xcolor}
\usepackage{orcidlink}
\usepackage{epsfig}
\usepackage{caption}
\usepackage{subcaption}
\usepackage{commath}
\hypersetup{colorlinks,citecolor=blue}
\hypersetup{colorlinks=true,linkcolor=magenta,filecolor=magenta,    urlcolor=blue}

\def\be{\begin{equation}}
\def\ee{\end{equation}}
\def\bea{\begin{eqnarray}}
\def\eea{\end{eqnarray}}

\begin{document}

\title{\bf Yano-Schrödinger Hyperfluid: Cosmological Implications}
\author{Himanshu Chaudhary}
\email{himanshu.chaudhary@ubbcluj.ro,\\
himanshuch1729@gmail.com}
\affiliation{Department of Physics, Babes-Bolyai University, Kogalniceanu Street, Cluj-Napoca 400084, Romania,}
\affiliation{Research Center of Astrophysics and Cosmology, Khazar University, Baku, AZ1096, 41 Mehseti Street, Azerbaijan}
\author{Saddam Hussain}
\email{saddamh@zjut.edu.cn}
\affiliation{Institute for Theoretical Physics and Cosmology, Zhejiang University of Technology, Hangzhou 310023, China}
\begin{abstract}
Perfect cosmological hyperfluids generalize the concept of a perfect fluid within the framework of metric affine gravity. These hyperfluids encode the microstructure of matter including shear, dilation, and spin via the hypermomentum tensor. In this paper, we focus on the observational constraints of the recently introduced Yano-Schrödinger hyperfluid, which sources a special type of nonmetricity, that preserves the lengths of vectors under autoparallel transport. We propose a model in which the effective nonmetricity contributions to pressure and matter density are related linearly as \( p_{\text{eff}} = \omega \rho_{\text{eff}} \). This assumption allows for a straightforward parameterization of deviations from standard cosmological behavior while maintaining analytical tractability. To constrain the effective equation of state parameter \( \omega \), we perform a Bayesian parameter estimation using Nested Sampling, implemented via the \texttt{PyPolyChord} library. We use Baryon Acoustic Oscillation measurements from the Dark Energy Spectroscopic Instrument (DESI) Data Release 2 (DR2), along with Type Ia supernova and Cosmic Chronometer data. In our analysis, we treat \( r_d \) as a free parameter, enabling late-time data to extract posterior distributions for the Hubble constant (\( H_0 \)) and the sound horizon (\( r_d \)), along with the corresponding model parameters. Our results yield \( H_0 = 67.4 \pm 4.0 \) km s\(^{-1}\) Mpc\(^{-1}\) and \( r_d = 148.8 \pm 7.4 \) Mpc, with \( \omega = -0.488 \). Finally, we use the logarithm of the Bayes factor to compare different Yano-Schrödinger model against the \(\Lambda\)CDM model. We find that the LESC model provides a better fit to the data, suggesting that modifications to metric-affine gravity could offer viable alternatives to the standard cosmological paradigm.
\end{abstract}

\maketitle
\tableofcontents

\section{Introduction}\label{sec_1}
The $\Lambda$CDM model has been highly successful in explaining many aspects of modern cosmology. It describes the Universe as being mainly composed of cold dark matter (CDM) and a cosmological constant ($\Lambda$), which drives cosmic acceleration. Precise observations have allowed scientists to measure key parameters of this model with great accuracy, particularly the Hubble constant ($H_0$), which represents the Universe’s expansion rate. However, different methods of measuring $H_0$ have led to a significant discrepancy known as the \textit{Hubble Tension}. Estimates based on early Universe observations, such as cosmic microwave background (CMB) data from the Planck satellite, suggest $H_0 = 67.4 \pm 0.5$ km/s/Mpc \cite{collaboration2020planck}. In contrast, local Universe measurements using the cosmic distance ladder calibrated with Cepheid variables and Type Ia supernovae yield a higher value, such as $H_0 = 73.04 \pm 1.04$ km/s/Mpc \cite{riess2022comprehensive}. This inconsistency, at a statistical significance of about 4$\sigma$ to 5.7$\sigma$, suggests a fundamental issue that cannot be easily explained by measurement errors alone.

This debate raises questions about the validity of the $\Lambda$CDM model. Potential explanations encompass systematic observational errors, calibration uncertainties, or the possibility of physics beyond the Standard Model. To explore this, alternative cosmological frameworks have been proposed \cite{miao2018h0,di2021investigating,li2019simple,zhao2017dynamical,ryan2019baryon,ding2020gigaparsec,liu2020can,vagnozzi2020new,poulin2019early,yang2019observational,xu2018detecting,di2016reconciling,huang2016dark,Hussain:2024jdt}. Additionally, independent techniques such as gravitational wave events \cite{mooley2018superluminal,ligo2017gravitational,hotokezaka2019hubble}, fast radio bursts (FRBs) \cite{wu20228,james2022measurement}, Megamaser \cite{pesce2020megamaser,reid2019improved,kuo2015megamaser}, Quasar Lensing \cite{millon2020tdcosmo,wong2020h0licow}, the red giant branch tip method (TRGs) \cite{freedman2019carnegie,freedman2020calibration,freedman2021measurements}, and Baryon Acoustic Oscillations (BAOs) \cite{addison2018elucidating} provide alternative ways to estimate $H_0$. For example, LIGO/Virgo's analysis of GW190412 combined with optical data from the Dark Energy Survey estimated $H_0 = 77.96^{+23.0}_{-5.03}$ km/s/Mpc \cite{ballard2023dark}, while data from the DELVE survey and LIGO/Virgo's first three runs suggest $H_0 = 68.84^{+15.51}_{-7.74}$ km/s/Mpc \cite{alfradique2024dark}. Another effective method for studying cosmic expansion is the analysis of Baryon Acoustic Oscillations (BAOs). These oscillations originated as sound waves traveling through the hot plasma of the early Universe and became imprinted in the large-scale distribution of galaxies after recombination. BAO measurements from surveys such as the Dark Energy Spectroscopic Instrument and the Dark Energy Survey are crucial for cosmological analysis. BAO observations provide key measurements, including \( D_A(z)/r_d \), \( D_V(z)/r_d \), \( D_M(z)/r_d \), \( D_H/r_d \), and \( H(z) \cdot r_d \), where \( r_d \) represents the comoving sound horizon at the baryon decoupling redshift \( z_d \). The Hubble constant \( H_0 \) and the sound horizon \( r_d \) are strongly linked, connecting early and late Universe measurements. The value of \( r_d \) is determined by early Universe physics and has been precisely constrained using Planck data \cite{collaboration2020planck}. An alternative approach to calibrating \( r_d \) involves using BAO data in combination with low-redshift observations while treating \( r_d \) as a free parameter. This model-independent technique avoids assumptions about early Universe conditions and the physics of recombination, offering an independent method to estimate cosmic expansion parameters \cite{pogosian2020recombination,jedamzik2021reducing,pogosian2024consistency,lin2021early,vagnozzi2023seven,Hussain:2024qrd}.

In parallel, alternative gravity theories have been explored to address unresolved issues in Einstein’s General Relativity (GR) and to explain cosmic phenomena that GR struggles with. The most relevant theoretical challenges in the concordance \(\Lambda\)CDM model involve the nature of dark energy and dark matter. These components are essential for accurately describing observational data in GR, yet their fundamental nature remains unknown. Furthermore, even with the inclusion of dark matter and dark energy, the \(\Lambda\)CDM model faces significant challenges, particularly in reconciling early-time and late-time predictions of cosmic expansion. To address these issues, theorists have proposed that while Einstein’s theory is highly successful on Solar System scales, it may require modifications at cosmological scales. Many alternative gravity theories exist, which can broadly be categorized into three main classes:
\begin{enumerate}
    \item Theories that extend the gravitational action, such as \(f(R), f(R,T), f(R,L_m)\) gravity \cite{RevModPhys.82.451, PhysRevD.84.024020, Harko2010}.
    \item Theories that modify the underlying geometry, such as metric-affine gravity (MAG) \cite{HEHL19951}.
    \item Theories that alter both the action and the geometry, such as \(f(Q), f(Q,T), f(\mathbb{T}), f(\mathbb{T},T)\) gravity \cite{HEISENBERG20241, Xu_2019, Cai_2016, Harko_2014}.
\end{enumerate}
The second class is particularly interesting, as it is deeply rooted in gauge-theoretic principles \cite{blagojević2022gaugetheoriesgravitation}. These have proven to be successfull for the description of elementary particle physics: the standard model is a $SU(3)\times SU(2) \times U(1)$ gauge theory. In metric-affine gravity, instead of taking the aforementioned group, one considers the Affine group, $GA(4,\mathbb{R})=T(4) \times GL(4,\mathbb{R})$ as the gauge group. In this approach, the metric and the connection are treated as independent variables: their dynamics are governed by the field equations obtained from variational principles. As such, torsion and nonmetricity are a consequence of the geometry-matter coupling, induced by the hypermomentum \cite{HehlKerlickHeyde+1976+111+114,Iosifidis_2021}.

Geometric modifications with torsion date back to the early work of Cartan \cite{Cartan1,Cartan2,Cartan3,Cartan4}, who was the first to introduce this concept into differential geometry. In geometries with torsion, the Ricci scalar loses its symmetry \cite{Riccicalculusbook}, necessitating a distinct form of matter on the right-hand side of the Einstein equations to source the antisymmetric component. In Einstein-Cartan theory, this is achieved by linking these antisymmetric terms to the spin of matter, such as through a Weysenhoff fluid \cite{YNObukhov_1987}. Torsion’s incorporation into cosmology was later explored by Kranas et al. \cite{Kranas_2019}, demonstrating that it may act as either the cosmological constant or spatial curvature, significantly affecting the Universe’s dynamics.
In \cite{CSILLAG2024101596}, the Friedmann equations, adjusted only by variable rescaling, are derived from a semi-symmetric (often termed vectorial) torsion framework \cite{FriedmannSchouten}. In \cite{Chaudhary2024}, the simplest torsion-based models are compared with observational data.

In contrast, theories incorporating nonmetricity within the metric-affine framework have received less attention. The earliest formulation stems from Weyl \cite{Weyl}, who aimed to unify electromagnetic and gravitational interactions. We believe Einstein’s objection to the non-preservation of lengths under parallel transport likely contributed to the theory’s prolonged neglect. Recently, however, this geometry has been reexamined from new perspectives \cite{Gh1,Gh2,Harkodarkmatter,Harkoweylastro} and applied to semimetals \cite{palumbo2024weylgeometryweylsemimetals}. Schrödinger introduced a connection \cite{Schrodinger} that resolves Einstein’s critique of Weyl’s geometry by preserving lengths under autoparallel transport. This geometry has gained traction in recent works by Ravera et al. \cite{SilkeKlemm}, Ming et al. \cite{HarkoSchr}, and Csillag et al. \cite{Csillag_2024,Csillag_2024jcap}. The preservation of lengths under autoparallel transport enables a natural generalization of the Raychaudhuri and Sachs equations within this framework \cite{Csillag_2024}, a task complicated by the Weyl connection or other nonmetric connections \cite{agakine,Agashe_2024}. This difficulty arises because nonmetricity may alter the causal structure of timelike or null congruences, potentially transforming a null vector into a timelike one or a timelike vector into a null one. In Schrödinger’s geometry, however, timelike and null autoparallels maintain their causal nature \cite{Csillag_2024}. In \cite{CSILLAG2024101596}, it is demonstrated that symmetrizing a semi-symmetric torsion over the appropriate indices naturally yields a Schrödinger connection. Similarities between torsion and nonmetricity in the metric-affine framework are explored in \cite{Iosifidis_2019torsionduality}, where Weyl-type nonmetricity is shown to be equivalent to vectorial (or semi-symmetric) torsion under projective invariance.

As previously noted, perfect fluid models, such as the Weysenhoff fluid, have been explored within the metric-affine framework. A comprehensive generalization was proposed by Iosifidis \cite{Iosifidis_2020fluids,Iosifidis_2021hyperfluid}, introducing the \textit{perfect cosmological hyperfluid}. This model extends the conventional perfect fluid of GR by incorporating microscopic properties of matter—such as shear, spin, and dilation—which source spacetime geometry through the connection field equations. Despite growing interest in hyperfluid models \cite{Obukhov_2023,Iosifidis_2023jcap,Iosifidis_2024jcap,andrei2024friedmanncosmologyhyperfluids,Iosifidis_2024hyperhdro}, their compatibility with observational data remains underexplored in the literature. These models have primarily been examined theoretically. This gap motivates the present study, which investigates the compatibility of the recently developed Yano-Schrödinger hyperfluid \cite{Csillag_2024} with cosmological observational data.

The paper is structured as follows: In Section \ref{sec_1}, we provide an introduction to the topic. Section \ref{sec_2} reviews the theoretical foundations of the Yano–Schrödinger hyperfluid, including its geometric structure and the associated gravitational field equations. In Section \ref{sec_3}, we present a cosmological model based on a linear effective equation of state, $p_{\text{eff}} = \omega \rho_{\text{eff}}$, and describe the methodology and datasets used for Bayesian inference. We then perform a comparative analysis with the $\Lambda$CDM model using $H(z)$ and $\mu(z)$, followed by a cosmographic study and the $Om(z)$ diagnostic. We also present the evolution of the matter density function $r(z)$ and the nonmetricity function $\Psi(z)$. Section \ref{sec_4} presents the main results. Finally, in Section \ref{sec_5}, we summarize our conclusions and discuss possible directions for future work.
\section{The Yano-Schrödinger hyperfluid}\label{sec_2}
The Yano-Schrödinger hyperfluid model is a special case of the general theory of perfect hyperfluids. More precisely, the hypermomentum part of the gravitational action is designed such that the solutions of the connection field equations yield precisely the Yano-Schrödinger connection. Before we present the mathematical details, let us briefly review the Yano-Schrödinger geometry, which is a special case of non-metric geometry.

\subsection{Yano-Schrödinger geometry}
The most general connection in metric-affine geometry, which is described by torsion and nonmetricity takes the form
\begin{equation}\label{generalconnection}
\begin{aligned}
    \tensor{{\Gamma}}{^\mu _\nu _\rho}=\tensor{\gamma}{^\mu _\nu _\rho} &+ \frac{1}{2} g^{\lambda \mu}(-Q_{\lambda \nu \rho}+ Q_{\rho \lambda \nu} + Q_{\nu \rho \lambda})\\
    &- \frac{1}{2}g^{\lambda \mu}(T_{\rho \nu \lambda}+T_{\nu \rho \lambda}- T_{\lambda \rho \nu}),
    \end{aligned}
\end{equation}
where $\tensor{\gamma}{^\mu_\nu _\rho}$ denotes the Christoffel symbols of the Levi-Civita connection, the nonmetricity tensor $Q_{\lambda \nu \rho}$ measures the failure of preservation of lengths under parallel transport, and the torsion tensor $T_{\lambda \nu \rho}$ is the antisymmetric part of the connection coefficient functions.\\
The Yano-Schrödinger geometry is a special case of the above formulation, where torsion is set to zero and the nonmetricity is vectorial, and given by
\begin{equation}
    \overset{YS}{Q}_{\lambda \nu \rho}=-w_{\lambda} g_{\nu \rho}+\frac{1}{2} w_{\rho} g_{\lambda \nu} + \frac{1}{2}w_{\nu} g_{\rho \lambda},
\end{equation}
where $w_{\mu}$ is a one-form. Note that this is very similar to Weyl geometry, where nonmetricity takes the following form
\begin{equation}
    \overset{W}{Q}_{\lambda \nu \rho}=-w_{\lambda} g_{\nu \rho}.
\end{equation} 
The difference between the two nonmetricities is completely determined by the Weyl one-form $w$, and is given by
\begin{equation}
    \overset{W}{Q}_{\lambda \nu \rho}- \overset{YS}{Q}_{\lambda \nu \rho}=-\frac{1}{2} w_{\rho} g_{\lambda \nu} + \frac{1}{2}w_{\nu} g_{\rho \lambda}.
\end{equation}
This minor difference leads to a very physically desirable property, namely to the existence of fixed-length vectors. For a more detailed description, we refer the reader to \cite{Csillag_2024}. By substituting the form of the Yano-Schrödinger nonmetricity in equation \eqref{generalconnection}, we obtain
\begin{equation}\label{thisconnection}
    \tensor{\Gamma}{^\mu _\nu _\rho}=\tensor{\gamma}{^\mu _\nu _\rho}+ w^{\mu} g_{\nu \rho} - \frac{1}{2} w_{\rho} \delta^{\mu}_{\nu} - \frac{1}{2} w_{\nu} \delta^{\mu}_{\rho}.
\end{equation}
In \cite{Csillag_2024} the curvature tensors of this connection have been computed. The Ricci tensor reads
\begin{equation}
\begin{aligned}
    R_{\mu \nu}=\overset{\circ}{R}_{\mu \nu}+ g_{\mu \nu} \overset{\circ}{\nabla}_{\alpha} w^{\alpha} &- \frac{1}{2} \overset{\circ}{\nabla}_{\mu} w_{\nu} + \overset{\circ}{\nabla}_{\nu} w_{\mu}\\
    &- \frac{1}{2} g_{\mu \nu} w^\alpha w_\alpha -\frac{1}{4} w_\mu w_\nu,
\end{aligned}
\end{equation}
while for the Ricci scalar, one has
\begin{equation}
    R=\overset{\circ}{R}+\frac{9}{2} \overset{\circ}{\nabla}_{\mu} w^\mu - \frac{9}{4} w_\mu w^\mu.
\end{equation}
\subsection{The gravitational field equations}
The gravitational field equations of the proposed theory are derived using the Palatini formalism. Here, torsion is explicitly assumed to vanish from the outset. The general class of theories under consideration is defined by the action
\begin{equation}
    S=\frac{1}{\kappa} \int \sqrt{-g} R(\Gamma)+S_{M}(g,\Phi,\Gamma).
\end{equation}
The variation with respect to the metric and connection yield the field equations
\begin{equation}
    R_{(\mu \nu)}(\Gamma)-\frac{1}{2} g_{\mu \nu}R(\Gamma)=\kappa T_{\mu \nu}, \;\; \tensor{P}{_\lambda ^{(\mu \nu)}}=\kappa \tensor{\Delta}{_\lambda^{(\mu \nu)}},
\end{equation}
where we have the energy-momentum and hypermomentum sources
\begin{equation}
    T_{\mu \nu}=\frac{2}{\sqrt{-g}}\frac{\delta \left( \sqrt{-g} \mathcal{L}_{M}\right)}{\delta g^{\mu \nu}}, \;  \tensor{\Delta}{_\lambda^{\mu \nu}}=-\frac{2}{\sqrt{-g}} \frac{\delta \left( \sqrt{-g} \mathcal{L}_{M} \right)}{\delta \tensor{\Gamma}{^\lambda _\mu _\nu}},
\end{equation}
and the Palatini tensor
\begin{equation}
     \tensor{P}{_\lambda^{(\mu \nu)}}=\frac{1}{2} Q_{\lambda} g^{\mu \nu}- \tensor{Q}{_\lambda ^{\mu \nu}}+ \left(q^{(\mu}-\frac{1}{2} Q^{(\mu} \right) \tensor{\delta}{_\lambda ^{\nu )}},
\end{equation}
with  $Q_{\lambda}:=Q_{\lambda \mu \nu} g^{\mu \nu}$  and  $q_{\nu}:=Q_{\lambda \mu \nu}g^{\lambda \mu}$. For matter, we now take a hyperfluid $S_{M}=S_{hyp}$, whose action is described by
\begin{equation}
\begin{aligned}
    S_{hyp}=\int& \mathrm{d}^4x \Bigg[ J^{\mu} \left(\varphi_{,\mu}+s \theta_{,\mu} + \beta_{k} \alpha^{k}_{, \mu} \right) \\
    &-\frac{\sqrt{-g}}{2}\left( 2 \rho(n,s,D)-\frac{5}{2} Q_{\mu} D^{\mu}+q_{\mu} D^{\mu}\right) \Bigg],
\end{aligned}
\end{equation}
where 
\begin{equation}
    D^{\mu}=\frac{D}{2 \kappa} u^{\mu}, \; \; \text{for some smooth function} \; \; D, 
\end{equation}
and $J^{\mu}$, representing the particle flux density, depends on the fluid variables—number density $n$ and comoving velocity $u^{\mu}$—via:
\begin{multline}
	J^{\mu} = \sqrt{-g} \  n u^{\mu}, \quad |J| = \sqrt{-g_{\mu\nu} J^{\mu} J^{\nu}}, \quad n = \frac{|J|}{\sqrt{-g}}, \\ u^{\mu} u_{\mu} = -1.
\end{multline}
In the Lagrangian, the variables $(\varphi, \theta, \alpha^k, \beta^k)$\footnote{where the index $k$ ranges from 1 to 3.} serve as Lagrange multipliers, while $s$, the entropy per particle, was introduced by Brown \cite{Brown:1992kc} and later incorporated into actions for perfect fluids in various studies \cite{Boehmer:2015kta,Chatterjee:2021ijw,Hussain:2023kwk}. Upon varying the aforementioned action gives the hypermomentum tensor 
 \begin{equation}
     \Delta_{\lambda \mu \nu}=\frac{D}{2\kappa} \left[h_{\mu \nu} u_{\lambda} -4 h_{\lambda (\nu} u_{\mu)} +3 u_\mu u_\nu u_\lambda \right],
 \end{equation}
which precisely sources the Palatini tensor of the Yano-Schrödinger geometry \cite{Csillag_2024}.\\
It is also interesting to mention that the hypermomentum tensor of a generic torsion-free cosmological hyperfluid takes the form
\begin{equation}
    \Delta_{\lambda \mu \nu}=\omega u_\lambda u_\mu u_\nu + \psi u_\lambda h_{\mu \nu}+\phi u_\nu h_{\lambda \mu}+\chi u_\mu h_{\lambda \nu},
\end{equation}
where $\omega, \psi,\phi, \chi$ are smooth functions of time. For the Yano-Schrödinger hyperfluid, these are all described by a single function $D(t)$, but differ by multiplicative constants
\begin{equation}
    \phi(t)=\chi(t)=-\frac{D(t)}{\kappa}, \; \; \psi(t)=-\frac{D(t)}{2\kappa}, \; \; \omega(t)=\frac{3D(t)}{2\kappa}.
\end{equation}
The metric variation leads to the field equations \cite{Csillag_2024}
\begin{equation}
\begin{aligned}
    \overset{\circ}{R}_{\mu \nu} &- \frac{1}{2} g_{\mu \nu} \overset{\circ}{R} -\frac{5}{4} g_{\mu \nu} \overset{\circ}{\nabla}_{\alpha} w^{\alpha}+\frac{1}{4} \left(\overset{\circ}{\nabla}_{\mu} w_{\nu} + \overset{\circ}{\nabla}_{\nu} w_{\mu} \right)\\
    &+ \frac{5}{8} g_{\mu \nu} w^{\alpha} w_{\alpha}-\frac{1}{4} w_{\nu} w_{\mu}=8 \pi T_{\mu \nu}.
\end{aligned}
\end{equation}
Note that since the connection field equation is algebraic, this variational principle does not describe dynamics for the vector field $w$. This will be obtained later, by imposing an equation of state in a cosmological setting \cite{Csillag_2024}.
\section{Yano-Schrödinger hyperfluid Cosmology} \label{sec_3}
\subsection{Linear Effective Equation of State Cosmological (LESC) Model}
In this section, we review the cosmological evolution of a Yano-Schrödinger (YS) hyperfluid, and obtain a novel cosmological model, by imposing a linear effective equation of state. We work in the spatially flat FLRW metric
\begin{equation}
    ds^2=-dt^2+a^2(t) \delta_{ij} dx^i dx^j,
\end{equation}
where \(a(t)\) denotes the scale factor. The matter source is taken as a standard perfect fluid, described by the energy momentum tensor
\begin{equation}
    T_{\mu \nu}=\rho u_\mu u_\nu + p( u_\mu u_\nu+g_{\mu \nu}).
\end{equation}
In a comoving frame with $u_\mu=(1,0,0,0)$, the nonmetricity vector is characterized by a smooth function
\begin{equation}
    w_{\nu}=(\psi(t),0,0,0),
\end{equation}
in accordance with the cosmological principle. In these notations, the Friedmann equations read \cite{Csillag_2024}
\begin{equation}\label{Friedmann1}
    3H^2=8\pi \rho +\frac{3}{2} \dot \psi +\frac{15}{2} H \psi - \frac{9}{8} \psi^2=8\pi(\rho+\rho_{eff}),
\end{equation}
\begin{equation}\label{Friedmann2}
        3H^2+2 \dot{H}=-8\pi p +\frac{5}{2} \dot \psi +4 H \psi - \frac{3}{8} \psi^2=-8\pi(p+p_{eff}),
\end{equation}
where the Hubble parameter is
\begin{equation}
    H=\frac{\dot a(t)}{a(t)}.
\end{equation}
Additionally, we define dimensionless variables $(h,\tau,\Psi,r,P)$,  as follows:
\begin{equation}
	\label{dim}
    H=H_0h,\tau=H_0t,\psi=H_0 \Psi,\rho=\frac{3H_0^2}{8 \pi}r,p=\frac{3 H_0^2}{8 \pi} P.
\end{equation}
enabling the reformulation of the preceding equations into the form presented below:
\begin{equation}\label{Friedmannhyper}
    3h^2=3r+\frac{3}{2}\frac{d \Psi}{d \tau} + \frac{15}{2} h \Psi - \frac{9}{8} \Psi^2,
\end{equation}
\begin{equation}
    2\frac{dh}{d\tau}+3h^2=- 3P + \frac{5}{2}\frac{d\Psi}{d \tau}+4h \Psi - \frac{3}{8} \Psi^2.
\end{equation}
The Friedmann equations of a YS hyperfluid in redshift space become:
\begin{multline}\label{redshift1}
   h^2(z)=r(z)-\frac{1}{2} (1+z)h(z) \frac{d\Psi}{dz} +\frac{5}{2} h(z) \Psi(z) - \frac{3}{8} \Psi(z)^2,
\end{multline}
\begin{multline}\label{redshift2}
    3h^2(z)-2(1+z)h(z) \frac{dh(z)}{dz}=-3P(z)\\
    -\frac{5}{2}(1+z)h(z) \frac{d \Psi(z)}{dz}+4 h(z) \Psi(z) - \frac{3}{8} \Psi^2.
\end{multline}\\\\
We will consider the case of dust matter by setting \( P = 0 \) in Eqs \eqref{Friedmann1} and \eqref{Friedmann2}. Then, by imposing the condition \( p_{eff}=w \rho_{eff} \) and using the dimensional parameters defined in Eq~\eqref{dim}, one can get the Hubble function as a system of differential equations:
\begin{equation}
\frac{d\Psi(z)}{dz} = \frac{2 (8 + 15 \omega) h(z) \Psi(z) + \frac{13 (-1 + 3 \omega)}{4} \Psi^2(z)}{(5 + 3 \omega) (1 + z) h(z)},
\end{equation}
\begin{equation}
\begin{split}
\frac{dh(z)}{dz} &= \frac{1}{2 (1 + z) h(z)} \Bigg( (1 + z) h \frac{5}{2} \frac{d\Psi(z)}{dz} - 4 h(z) \Psi(z)\\ &+ \frac{3}{8} \Psi^2(z) 
\quad + 3 h^2(z) \Bigg).
\end{split}
\end{equation}
These equations have to be solved with the initial conditions $h(0)=1$ and $\Psi(0):=\Psi_0$.
\subsection{Methodology and Data Description}
To constrain the parameters of the LESC model in  hyperfluid framework, we adopt a Bayesian statistical approach. The model is governed by a system of coupled differential equations describing the evolution of the Hubble parameter \( H(z) \) as a function of redshift \( z \). These equations, derived from theoretical considerations, are solved numerically using the \texttt{solve\_ivp} function from the \texttt{scipy} library \cite{virtanen2020scipy}. For numerical integration, we employ the \texttt{Radau} method, which is well-suited for stiff differential equations, over the redshift range \( 0 \leq z \leq 3 \). To ensure accuracy, we set relative and absolute tolerances to \( 10^{-3} \) and \( 10^{-6} \), respectively. Once the numerical solutions are obtained, we construct a likelihood function to evaluate how well the model aligns with observational data. This function incorporates key datasets, including measurements from Cosmic Chronometers, the Pantheon$^{+}$ dataset (excluding SHOES calibration), and recent Baryon Acoustic Oscillation data from the Dark Energy Spectroscopic Instrument Year 2. 

To enforce physical constraints and incorporate prior knowledge, we treat the parameters \( H_0 \), \( \Psi_{0} \), \( \omega \), \( \mathcal{M} \), and \( r_d \) (Mpc) as free parameters, assuming the following uniform priors:
\begin{align*}
H_0 &\in [50, 100],         & \Psi_0 &\in [0, 1], \\
\omega &\in [-1, 0],        & \mathcal{M} &\in [-20, -18], \\
r_d &\in [100, 300]         &&
\end{align*}
To sample the posterior distribution, we use the Nested Sampling algorithm implemented via the \texttt{PyPolyChord} library\footnote{\url{https://github.com/PolyChord/PolyChordLite}} \cite{handley2015polychord}, which simultaneously estimates the Bayesian evidence and explores the posterior distribution. For this analysis, we set the number of live points to 300 and the sampling accuracy to 0.01 to ensure adequate convergence given the dimensionality of the parameter space.

The resulting posterior samples are used to compute credible intervals and visualize parameter constraints. We use the \texttt{getdist} library\footnote{\url{https://github.com/cmbant/getdist}} \cite{lewis2019getdist} to generate triangular plots, which effectively shows the marginal distributions and correlations between parameters. To determine the posterior distribution of the LESC model, we construct a likelihood function for each dataset. Below, we describe each dataset and the corresponding formulation of its likelihood function.
\begin{itemize}
    \item \textbf{Cosmic Chronometers:} In our analysis, we use a subset of 15 Hubble measurements from a total of 31 data points, spanning the redshift range \( 0.1791 \leq z \leq 1.965 \) \cite{moresco2012new,moresco2015raising,moresco20166}, obtained using the differential age technique \cite{jimenez2002constraining}. This method, based on passively evolving massive galaxies formed at \( z \sim 2-3 \), enables direct, model-independent estimation of the Hubble parameter via \( \Delta z / \Delta t \), minimizing astrophysical assumptions \cite{moresco2015raising,moresco20166}. To infer the parameter distributions, we utilize the likelihood function implemented in the GitLab repository\footnote{\url{https://gitlab.com/mmoresco/CCcovariance}}, which incorporates the full covariance matrix, thereby accounting for both statistical and systematic uncertainties \cite{moresco2018setting,moresco2020setting}.
    \item \textbf{Type Ia supernova:} We also use the Pantheon$^{+}$ without SHOES calibration, which comprises 1701 light curves from 1550 Type Ia Supernovae (SNe Ia) across a redshift range of \( 0.01 \leq z \leq 2.26 \) \cite{brout2022pantheon}. We utilize the Pantheon$^{+}$ only cosmosis likelihood defined in \cite{riess1998observational,astier2006supernova}\footnote{\url{https://github.com/PantheonPlusSH0ES/DataRelease}}. This approach incorporates both statistical and systematic uncertainties through a covariance matrix \cite{conley2010supernova}. The likelihood function is defined as follows: $\mathcal{L_{\text{SNe Ia}}} = {e}^{\frac{-1}{2}(\Delta \mathbf{D}^T \mathbf{C}^{-1}_{\text{total}} \Delta \mathbf{D})},$ where \(\Delta \mathbf{D}\) represents the vector of residuals between the observed distance moduli \(\mu(z_i)\) and the model-predicted distance moduli \(\mu_{\text{model}}(z_i, \theta)\). Each residual, \(\Delta D_i\), is computed as: $\Delta D_i = \mu(z_i) - \mu_{\text{model}}(z_i, \theta).$ The total covariance matrix, \(\mathbf{C}_{\text{total}}\), combines both statistical (\(\mathbf{C}_{\text{stat}}\)) and systematic (\(\mathbf{C}_{\text{sys}}\)) uncertainties. Its inverse, \(\mathbf{C}^{-1}_{\text{total}}\), is used to account for these uncertainties in the analysis. The model-predicted distance moduli are given by: $\mu_{\text{model}}(z_i) = 5 \log_{10} \left( \frac{d_L(z)}{\text{Mpc}} \right) + \mathcal{M} + 25,$ where the luminosity distance \( d_L(z) \) in a flat FLRW Universe is defined as: $d_L(z) = c(1 + z) \int_0^z \frac{dz'}{H(z')}.$ Here, \( c \) is the speed of light, and \( H(z) \) denotes the Hubble parameter. This method reveals a degeneracy between the parameters $\mathcal{M}$ and $H_0$. Consequently, external datasets are incorporated to resolve this degeneracy. 
    \item \textbf{Baryon Acoustic Oscillation:} We incorporate the latest set of 13 Baryon Acoustic Oscillation (BAO) measurements from the Dark Energy Spectroscopic Instrument (DESI) Data Release 2 (DR2) \cite{karim2025desi}, covering the redshift range $0.295 < z < 2.330$. These measurements are derived from a variety of tracers, including the Bright Galaxy Sample (BGS), Luminous Red Galaxies (LRG1, LRG2, LRG3), Emission Line Galaxies (ELG1 and ELG2), Quasars (QSO), and Lyman-$\alpha$ forest data\footnote{\url{https://github.com/CobayaSampler/bao_data}}. The BAO measurements are reported in terms of the Hubble distance $D_H(z)$, the comoving angular diameter distance $D_M(z)$, and the volume-averaged distance $D_V(z)$. To constrain cosmological parameters, we utilize the following dimensionless ratios: $D_M(z)/r_d$, $D_H(z)/r_d$, and $D_V(z)/r_d$, where $r_d$ is the comoving sound horizon at the drag epoch $z_d \approx 1060$. The sound horizon $r_d$ is computed from the integral: $r_d = \int_{z_d}^{\infty} \frac{c_s(z)}{H(z)} \, dz,$ where the sound speed $c_s(z)$ in the photon-baryon fluid is given by: $c_s(z) \approx \frac{c}{\sqrt{3 + \frac{9 \rho_b(z)}{4 \rho_\gamma(z)}}},$ with $\rho_b(z)$ and $\rho_\gamma(z)$ representing the baryon and photon energy densities, respectively. The normalized Hubble function $E(z) = H(z)/H_0$ depends on the cosmological model. Under the standard flat $\Lambda$CDM model, the Planck 2018 results yield an estimate of $r_d = 147.09 \pm 0.26 \, \text{Mpc}$ \cite{collaboration2020planck}. However, in our analysis, we treat $r_d$ as a free parameter, enabling late-time observational data to constrain model parameters. \cite{pogosian2020recombination,jedamzik2021reducing,pogosian2024consistency,lin2021early,vagnozzi2023seven}.
\end{itemize}    
The parameter distributions of the LESC model within the YS Hyperfluid framework are obtained by maximizing the likelihood function, \(\mathcal{L}\). The total likelihood function, denoted as $\mathcal{L}_{\text{Tot}}$, is expressed as: $\mathcal{L}_{\text{Tot}} = e^{\frac{-\chi^2_{\text{Tot}}}{2}}$, where 
\begin{equation}
	\chi_{\rm Tot}^2 = \chi_{\rm CC}^2 + \chi_{\rm SNe Ia}^{2} + \chi_{\rm BAO}^{2}\ .
\end{equation}
To compare the LESC model with the standard $\Lambda$CDM model, we use the Bayes factor $B_{ij}$, a statistical tool that quantifies the relative support of two competing models $M_i$ and $M_j$ based on observed data $d$. It is defined as: $B_{ij} = \frac{p(d \mid M_i)}{p(d \mid M_j)}$ where $p(d \mid M_i)$ and $p(d \mid M_j)$ are the Bayesian evidences (or marginal likelihoods) for models $M_i$ and $M_j$. Bayesian evidence accounts for both model fit and complexity, offering a balanced basis for comparison. Taking the natural logarithm gives a more interpretable form: $\ln(B_{ij}) = \ln p(d \mid M_i) - \ln p(d \mid M_j)$ In our analysis, \texttt{PolyChord} is used to compute $\log Z$ values numerically for each model. To interpret the strength of the evidence, we use Jeffreys' scale \cite{jeffreys1961theory}:
\begin{itemize}
\item $\ln(B_{ij}) < 1$: Inconclusive
\item $1 \leq \ln(B_{ij}) < 2.5$: Weak evidence
\item $2.5 \leq \ln(B_{ij}) < 5$: Moderate evidence
\item $\ln(B_{ij}) \geq 5$: Strong evidence
\end{itemize}
In our case, $B_{ij} = \frac{p(d \mid M_i)}{p(d \mid M_j)}$, model $M_i$ corresponds to the LESC model and model $M_j$ corresponds to the $\Lambda$CDM model. The term $p(d \mid M_i)$ represents the integrated likelihood of observing the data $d$ given the LESC model, averaged over the prior distribution of its parameters.
\begin{figure}
\begin{subfigure}{.46\textwidth}
\includegraphics[width=\linewidth]{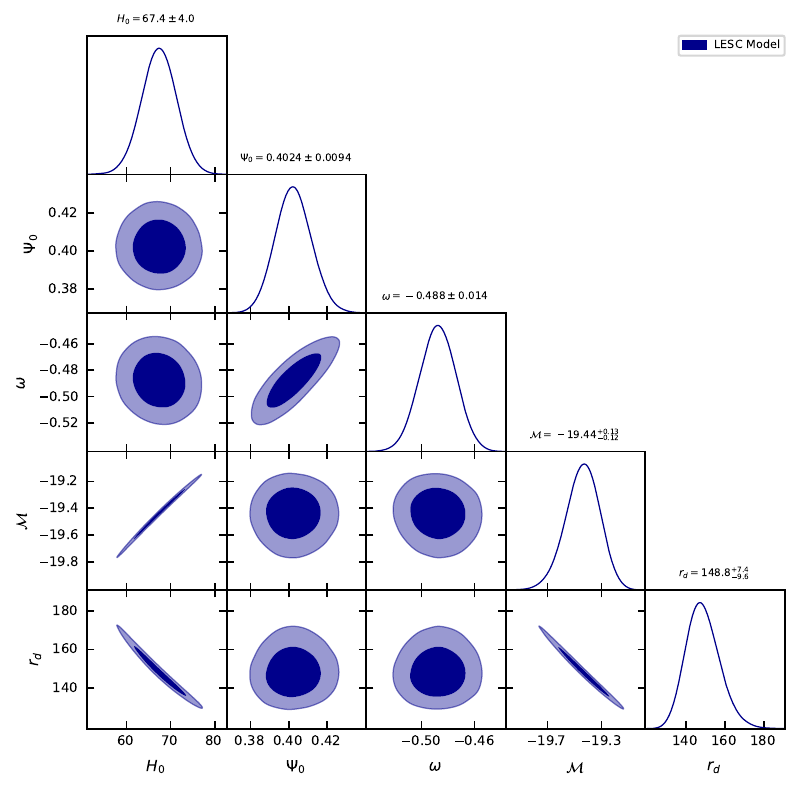}
\end{subfigure}
\caption{The posterior distributions of LESC model parameters in the YS hyperfluid framework at 68\% (1\(\sigma\)) and 95\% (2\(\sigma\)) credible intervals.}\label{fig_1}
\end{figure}
\begin{table*}
\resizebox{\textwidth}{!}{%
\begin{tabular}{lcccccccc}
\toprule
\textbf{Model} & $H_0$ [$\mathrm{km\,s^{-1}\,Mpc^{-1}}$] & $\Omega_{m0}$ & $\Psi_{0}$ & $\omega$ & $\mathcal{M}$ & $r_d$ [Mpc] & $\ln{B}$ & $\chi^{2}_{min}$ \\
\midrule
\textbf{Flat $\Lambda$CDM} & & & & \\
DESI DR2 + PP$^+$ + CC & $68.4 \pm 3.9$ & $0.3023 \pm 0.0086$ & & & $-19.42 \pm 0.12$ & $148.0 \pm 7.3$ & 0 & 1578.92\\
\addlinespace[0.2cm]
\textbf{LESC} & & & & \\
DESI DR2 + PP$^+$ + CC & $67.4 \pm 4.0$ & --- & $0.4024 \pm 0.0094$ & $-0.488 \pm 0.014$ & $-19.44 \pm 0.12$ & $148.8 \pm 7.4$ & 3.8235 & 1553.79 \\
\bottomrule
\end{tabular}
}
\caption{Best-fit parameter values with 68\% (1$\sigma$) credible intervals, including prior ranges, for the standard $\Lambda$CDM model compared with the LESC model within the YS hyperfluid framework.}\label{tab_1}
\end{table*}
\subsubsection{Comparative analysis with the $\Lambda$CDM model using $H(z)$ and $\mu(z)$}
After determining the mean values of the free parameters in the LESC model, it is essential to assess the model’s predictions in comparison to well-established \(\Lambda\)CDM model, which serves as a benchmark. A key aspect of this analysis involves studying the evolution of the Hubble parameter \(H(z)\). For the standard \(\Lambda\)CDM framework, the Hubble parameter is given by: $H(z) = H_0 \sqrt{\Omega_{m0} (1 + z)^3 + (1-\Omega_{m0})}.$ Here, we adopt the best-fit values $H_0 = 68.4 \, \mathrm{km\,s^{-1}\,Mpc^{-1}}$ and $\Omega_{m0} = 0.302$, obtained from the MCMC analysis using the combined observational dataset. The behavior of \(H(z)\) as a function of redshift is then examined for both the LESC model in the YS hyperfluid framework and the \(\Lambda\)CDM model, and the results are compared with the CC dataset. Additionally, we calculate the distance modulus \(\mu(z)\) to further evaluate the LESC model’s predictions. The distance modulus is defined as: $\mu(z) = 5 \log_{10}(D_L(z)) + 25,$
where \(D_L(z)\) is the luminosity distance. The luminosity distance itself is expressed as \(D_L(z) = (1 + z) \int_0^z \frac{c}{H(z')} \, dz'\). Here, \(c\) represents the speed of light in a vacuum, and \(H(z')\) denotes the Hubble parameter at redshift \(z'\). Using the best fit values obtained from the MCMC analysis, we compute the distance modulus for the LESC model, denoted as \(\mu_{\text{LESC}}(z)\), and compare it with the \(\Lambda\)CDM model, \(\mu_{\Lambda \text{CDM}}(z)\). Finally, these theoretical predictions are plotted alongside observational data from 1701 Type Ia supernovae (SNe Ia) to assess the model’s consistency with empirical measurements.  
\begin{figure*}
\begin{subfigure}{.46\textwidth}
\includegraphics[width=\linewidth]{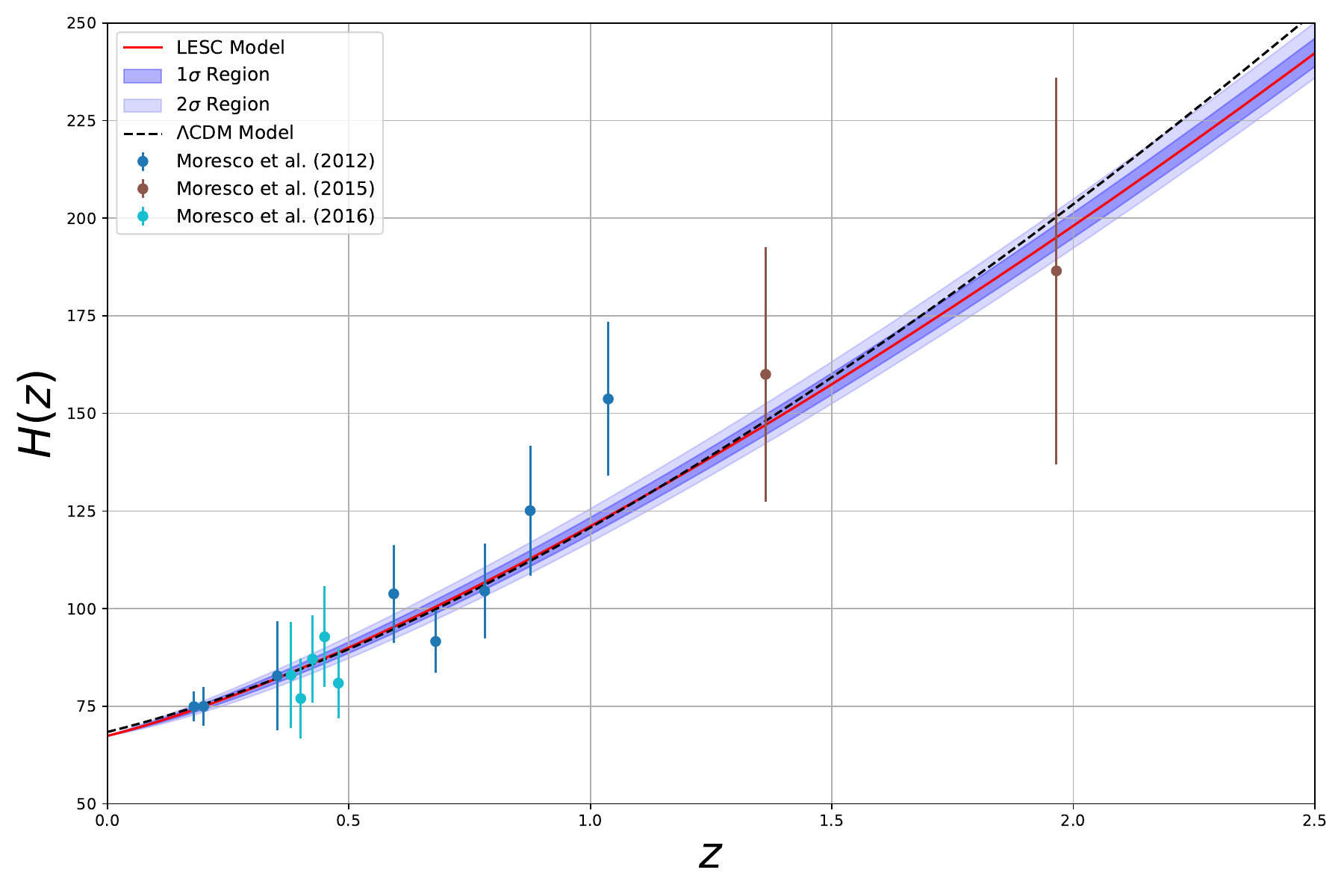}
\end{subfigure}
\hfil
\begin{subfigure}{.50\textwidth}
\includegraphics[width=\linewidth]{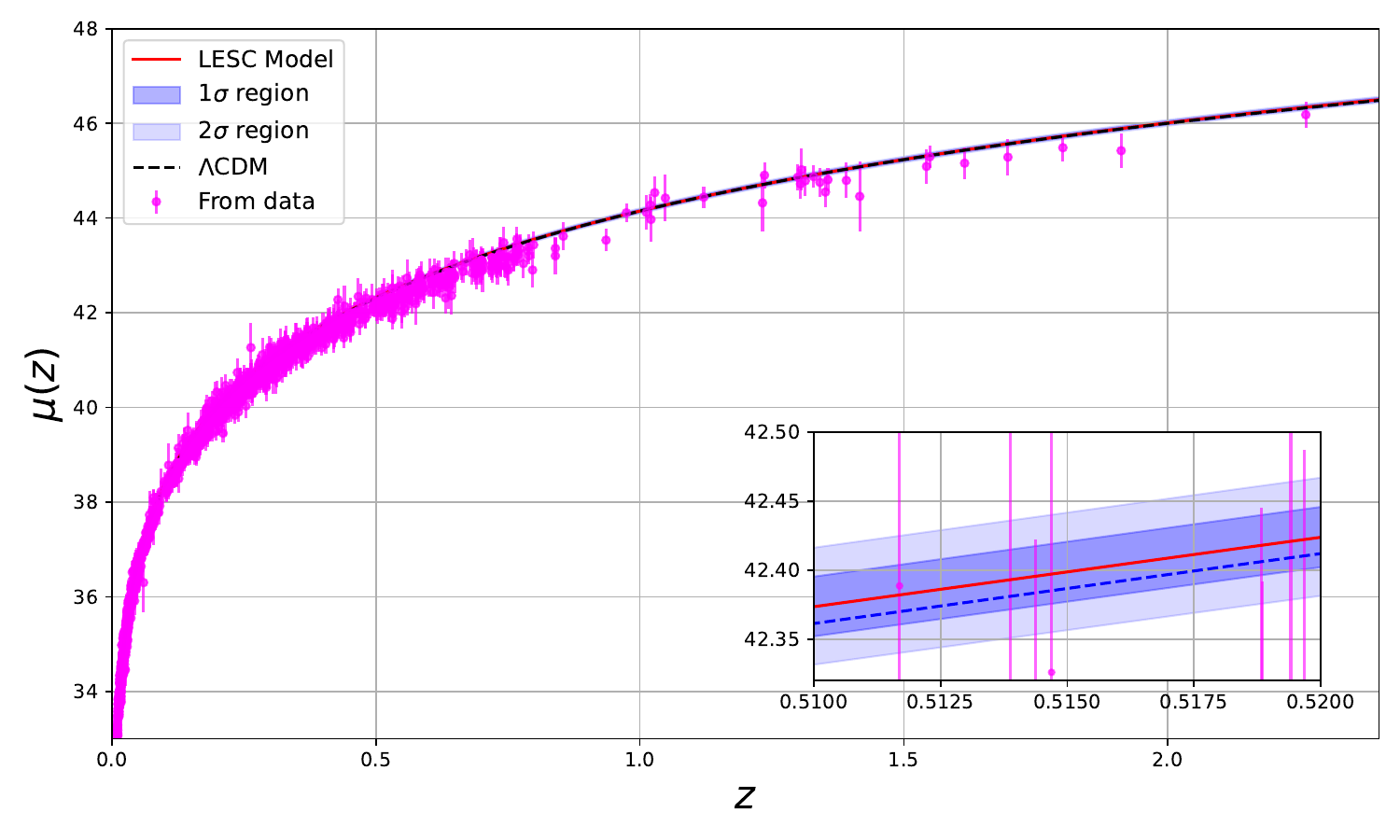}
\end{subfigure}
\hfil
\caption{This figure shows the $\Lambda$CDM and LESC models. The left panel shows the Hubble parameter $H(z)$ with CC measurements \cite{moresco2012new,moresco2015raising,moresco20166}, including the LESC prediction (red line), 1$\sigma$ and 2$\sigma$ confidence regions (shaded), and $\Lambda$CDM model (dashed line). The right panel displays the distance modulus $\mu(z)$ compared to Pantheon$^{+}$ measurements, with an inset highlighting model differences over a narrow redshift range.}\label{fig_2}
\end{figure*}
\subsection{Cosmographic analysis}
In this section, we analyze the expansion history of the universe using the deceleration parameter \( q(z) \) and the jerk parameter \( j(z) \), which provide insights into cosmic evolution \cite{aghanim2020planck,visser2004jerk}. The deceleration parameter determines whether the expansion is accelerating or decelerating, while the jerk parameter characterizes variations in acceleration. A key feature of the standard \(\Lambda\)CDM model is that the jerk parameter remains constant at \( j(z) = 1 \) \cite{visser2004jerk}. Using Planck 2018 parameters, the present-day deceleration parameter is approximately \( q(0) \approx -0.526 \), indicating that the Universe is currently undergoing accelerated expansion. Our approach involves comparing different cosmological models, specifically the LESC model, against \(\Lambda\)CDM. By analyzing the behavior of \( q(z) \) and \( j(z) \) across models, we aim to evaluate how well the LESC framework aligns with observational data and whether it provides a viable alternative in describing the Universe’s expansion \cite{riess1998observational,perlmutter1999measurements}.\\\\
\begin{figure*}
\begin{subfigure}{.49\textwidth}
\includegraphics[width=\linewidth]{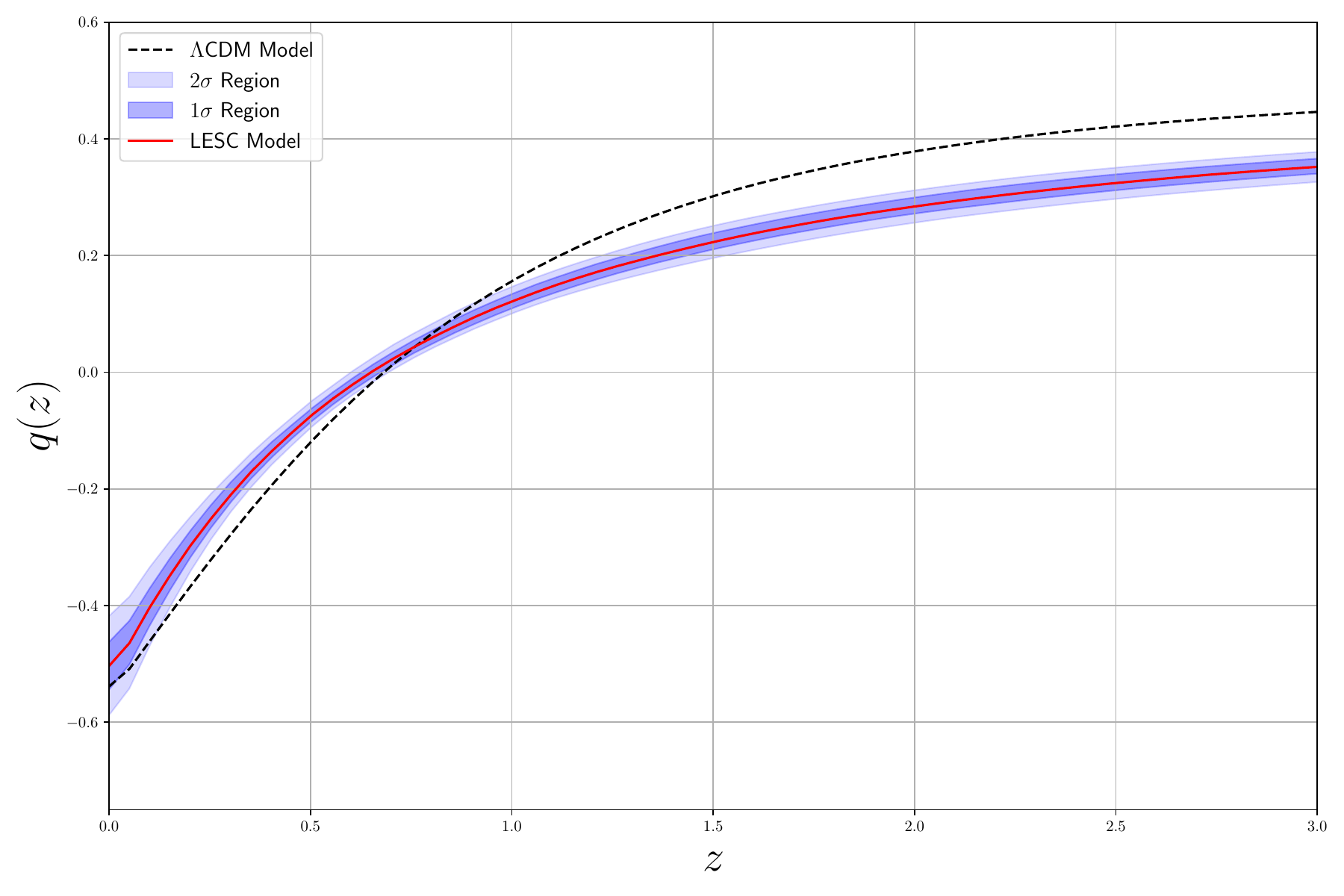}
\end{subfigure}
\hfil
\begin{subfigure}{.49\textwidth}
\includegraphics[width=\linewidth]{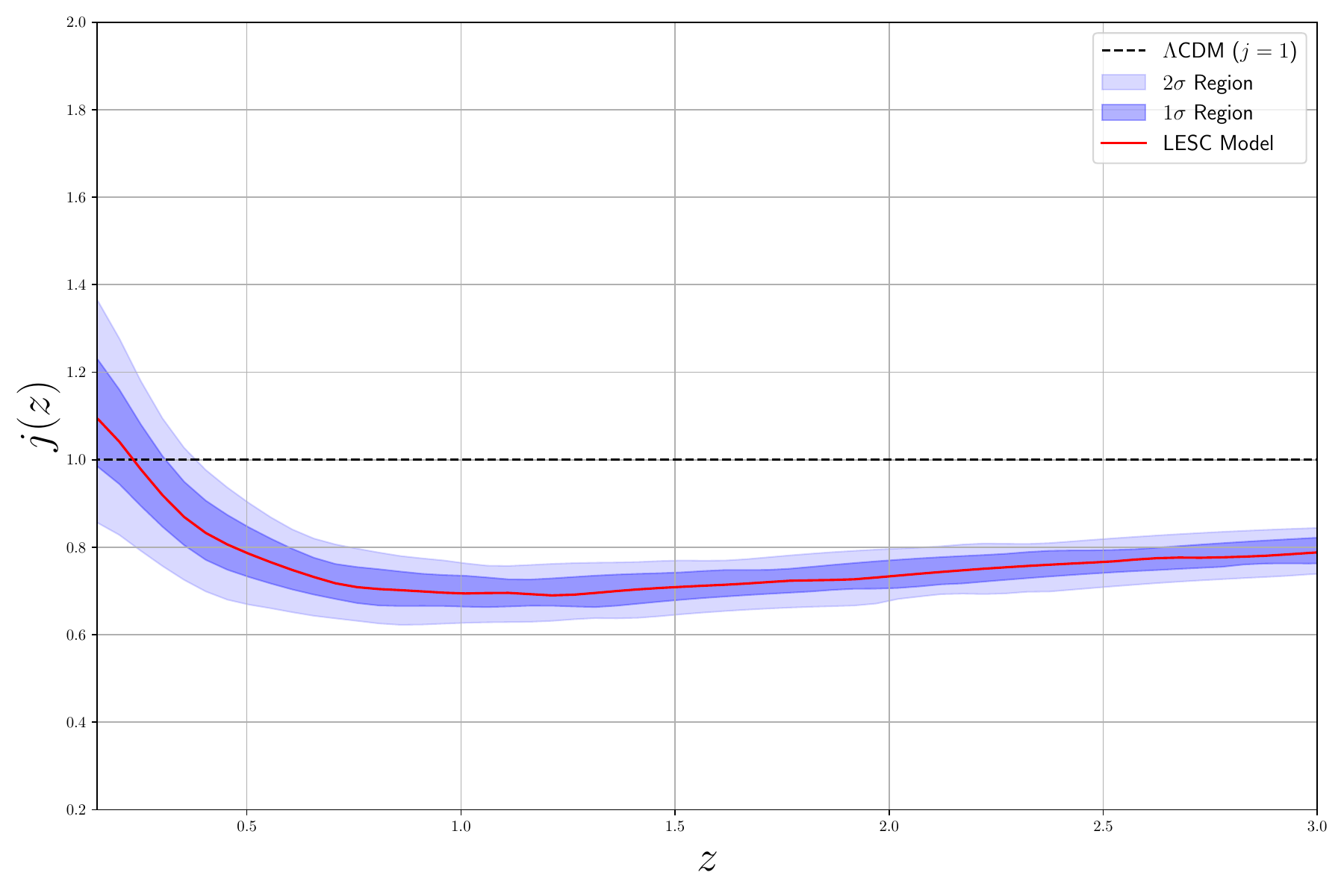}
\end{subfigure}
\hfil
\caption{This figure shows the evolution of the deceleration and jerk parameters with redshift for the $\Lambda$CDM and LESC models. The left panel shows the deceleration parameter $q(z)$, with the LESC model (red line), $1\sigma$ and $2\sigma$ confidence bands (shaded), and $\Lambda$CDM reference (dashed line). The right panel shows the jerk parameter $j(z)$, highlighting deviations from the $\Lambda$CDM prediction $j = 1$.}\label{fig_3}
\end{figure*}
\subsection{$Om(z)$ diagnostic}
To compare LESC model in YS hyperfluid framework with standard $\Lambda$CDM model, we'll also utilize the $Om(z)$ diagnostic \cite{Sahni,shahalam2015dynamics}, which is a crucial tool for differentiating alternative cosmological models. The $Om(z)$ function is defined as:
\begin{equation}
Om(z)=\frac{H^2(z)/H_0^2-1}{(1+z)^3-1}=\frac{h^2(z)-1}{(1+z)^3-1}.
\end{equation}
In the case of the $\Lambda$CDM model, $Om(z)$ is a constant equal to the present-day matter density, denoted as $r(0)=0.3166$. However, in other theories of gravity that deviate from the $\Lambda$CDM model, changes in the value of $Om(z)$ over time indicate different types of cosmic evolution. Specifically, if $Om(z)$ increases (positive slope), it suggests a phantom-like evolution. Conversely, if $Om(z)$ decreases (negative slope), it points to quintessence-like dynamics.
\begin{figure}
\centering
\includegraphics[scale=0.28]{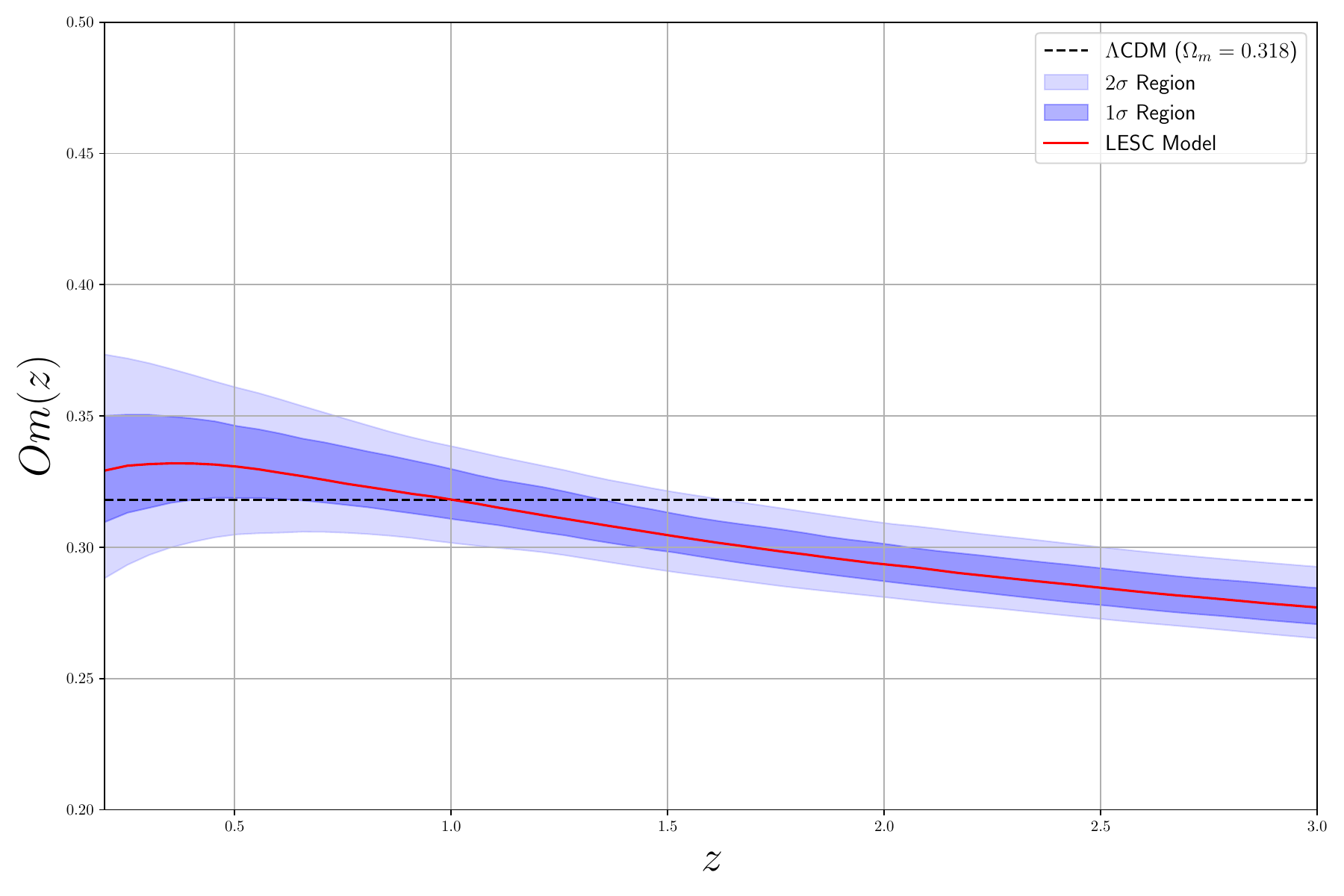}
\caption{This figure shows the evolution of the $Om(z)$ diagnostic comparing the $\Lambda$CDM and LESC models. The LESC prediction (red) includes $1\sigma$ and $2\sigma$ bands, with $\Lambda$CDM shown as a black dashed line.}\label{fig_4}
\end{figure}
\subsection{Matter density $r(z)$ and Nonmetricity \(\Psi(z)\)}
In this section, we will examine the behavior of two important quantities: matter density \( r(z) \), which describes the evolution of matter energy density with redshift \( z \), and nonmetricity \( \Psi(z) \) in the YS hyperfluid framework, which quantifies deviations from Levi-Civita connections and affects gravitational dynamics.
\begin{figure}
\centering
\includegraphics[scale=0.28]{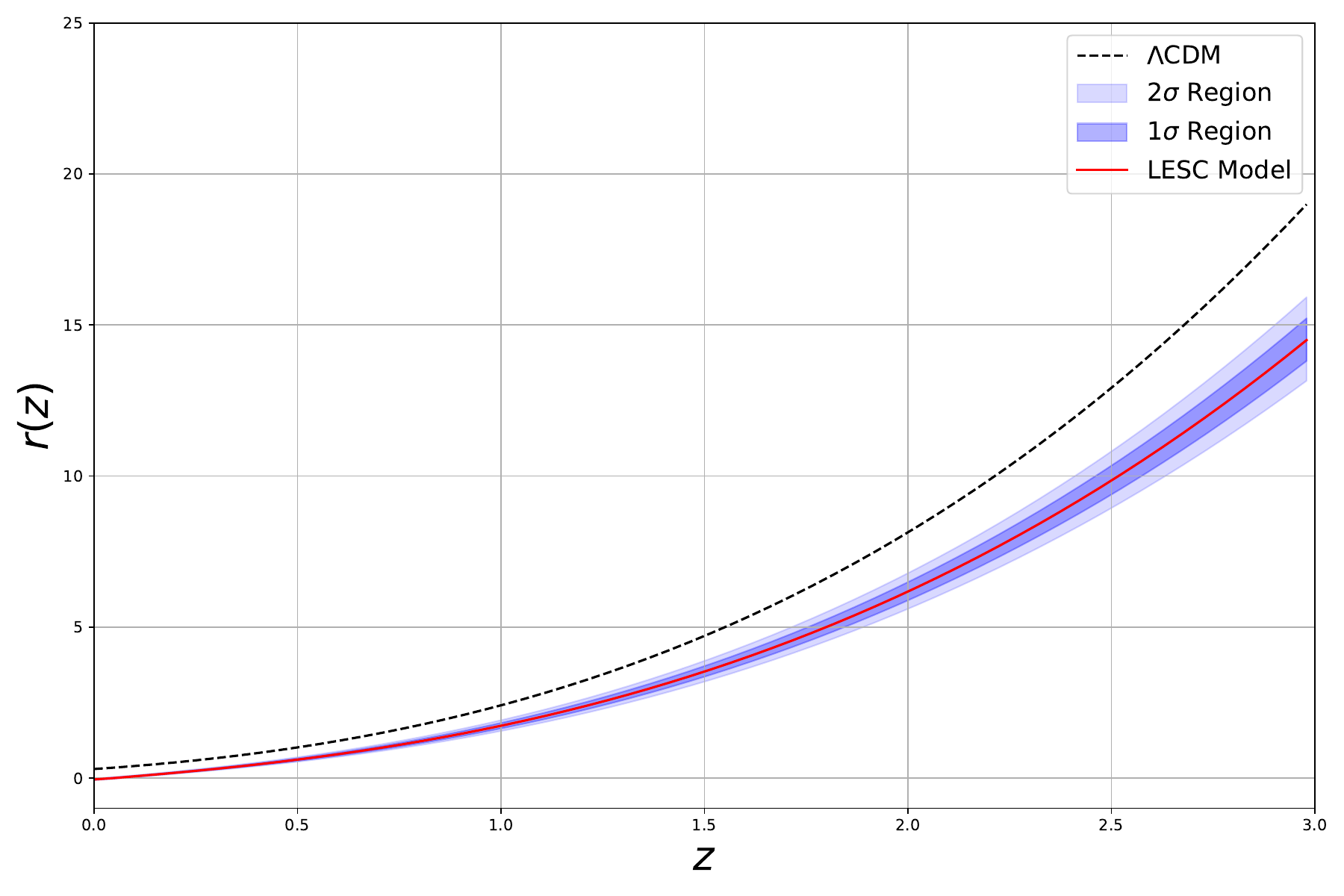}
\caption{This figure shows the evolution of the dimensionless matter density $r(z)$ for the $\Lambda$CDM and LESC models. The LESC prediction (red line) includes $1\sigma$ and $2\sigma$ confidence bands, while the $\Lambda$CDM model is shown as a black dashed line.}\label{fig_5}
\end{figure}
\begin{figure}
\centering
\includegraphics[scale=0.28]{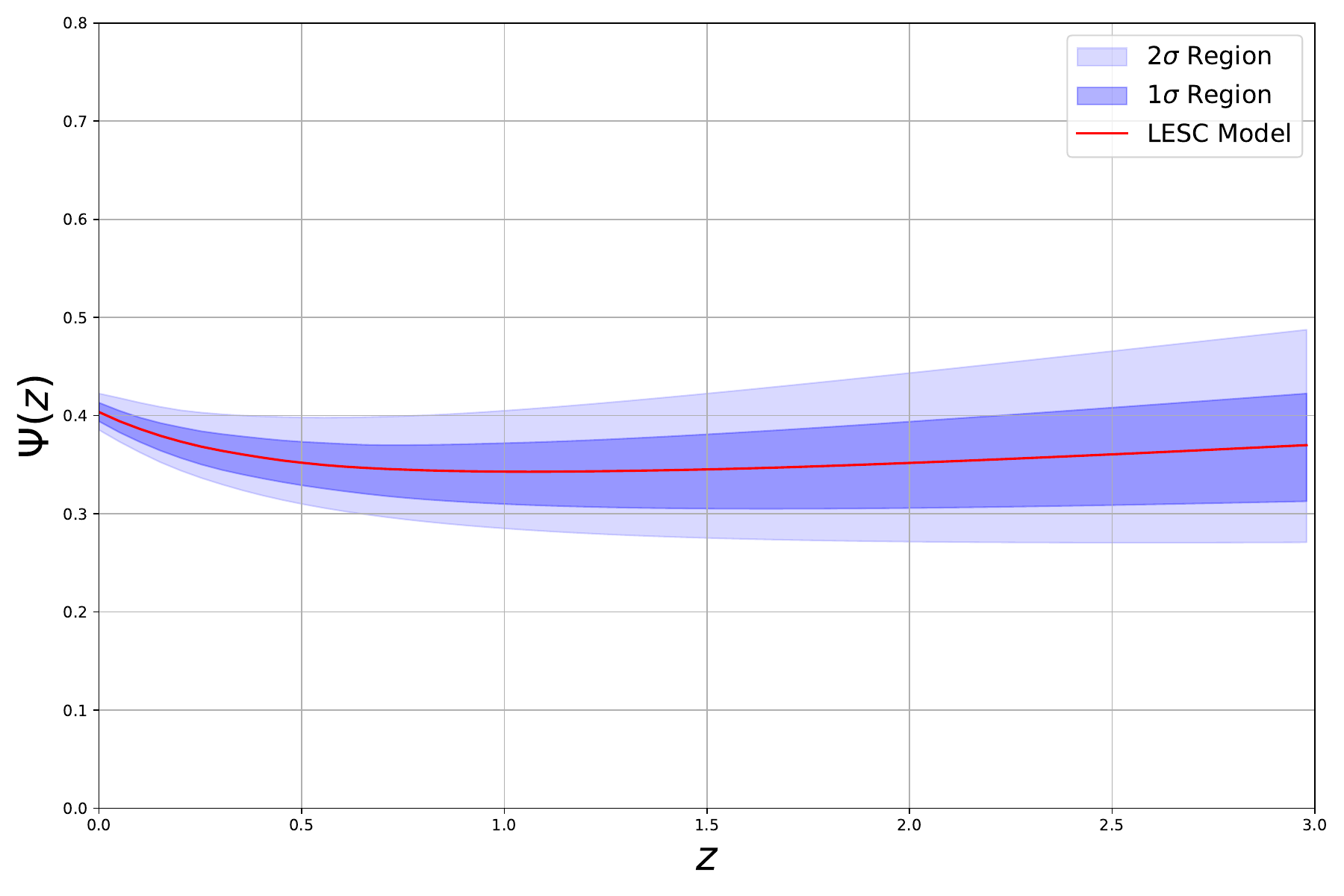}
\caption{This figure shows the evolution of the dimensionless nonmetricity function \(\Psi(z)\) in the LESC model. The red line shows the mean prediction, with shaded regions indicating the \(1\sigma\) and \(2\sigma\) confidence intervals.}
\label{fig_metrcitiy}
\end{figure}
\section{Results}\label{sec_4}
Fig. \ref{fig_1}, shows the corner plot shows the parameter constraints for the LESC model within the YS hyperfluid framework. The plot features 1D marginalized distributions along the diagonal and 2D contour plots in the off-diagonal terms, highlighting the correlations between different parameter pairs. Table~\ref{tab_1}, presents the mean values along with the $68$\% (1\(\sigma\)) credible intervals and prior ranges for both the \(\Lambda\)CDM model and the LESC model within the YS hyperfluid framework.

We observe that the extracted value of $H_0$ in both the $\Lambda$CDM and LESC models is consistent with the CC measurement from Moresco et al., when full systematics are taken into account. Additionally, the predicted value of the sound horizon $r_d$ in our analysis is higher than the value reported by Planck with high precision. This deviation, along with the broader uncertainties in both $H_0$ and $r_d$, can be attributed to the inclusion of the full covariance matrix in the CC dataset. The matrix incorporates systematic effects such as uncertainties in stellar metallicity (influenced by possible residual young stellar populations), variations in star formation history, assumptions about the initial mass function (IMF), the choice of stellar population synthesis models, and the use of different stellar libraries \cite{moresco2018setting}. On the other hand, the predicted values of \(\Omega_{m0}\) and \(\Omega_{\Lambda 0}\) are close to the values predicted by the Planck collaboration (\(\Omega_m = 0.315 \pm 0.007\), \(\Omega_\Lambda = 0.685 \pm 0.007\)).

Fig. \ref{fig_2}, shows the evolution of the Hubble parameter \( H(z) \) and the distance modulus \( \mu(z) \) for the \(\Lambda\)CDM and LESC models, compared against CC and SNe Ia measurements as a function of redshift. As shown in the left panel, the LESC model closely follows the standard $\Lambda$CDM prediction for the Hubble parameter $H(z)$ at low redshifts ($z < 1.5$), with the \(\Lambda\)CDM curve lying well within the 1$\sigma$ confidence region of the LESC model. At higher redshifts, a mild deviation emerges, although \(\Lambda\)CDM still remains within the 2$\sigma$ band, indicating no significant statistical tension. A similar trend is observed in the right panel for the distance modulus $\mu(z)$, where the predictions of both models are nearly indistinguishable across most of the redshift range. To better visualize the subtle differences, we include an inset plot highlighting a narrow redshift window. As the \(\Lambda\)CDM curve remains within the 1$\sigma$ region.

Fig~\ref{fig_3} shows the evolution of cosmographic parameters as a function of redshift. In the left panel, we present the deceleration parameter $q(z)$. At higher redshifts ($z > 0.5$), the $\Lambda$CDM model lies outside both the 1$\sigma$ and 2$\sigma$ confidence regions of the LESC prediction, indicating a significant statistical tension between the two models. In the right panel, which shows the jerk parameter $j(z)$, the $\Lambda$CDM value $j = 1$ also remains clearly outside both the 1$\sigma$ and 2$\sigma$ regions across the full redshift range. This implies a consistent and statistically notable deviation between the predictions of the LESC and $\Lambda$CDM models in terms of higher-order cosmographic behavior. However, it is worth noting that at the present epoch ($z = 0$), the $\Lambda$CDM model lies within the 1$\sigma$ region of the LESC model for both $q(z)$ and $j(z)$, indicating agreement at low redshift.

Fig.~\ref{fig_4} shows the evolution of the $Om(z)$ profile for the $\Lambda$CDM model and the LESC model within the YS hyperfluid framework. The $\Lambda$CDM model remains constant by definition, while the LESC model exhibits a gradual, monotonic decrease with redshift. This behavior is characteristic of quintessence-like evolution. At low redshift, the $\Lambda$CDM curve lies within the 1$\sigma$ confidence region of the LESC model, indicating consistency between the two models at the statistical level. However, as redshift increases, the $\Lambda$CDM prediction moves outside the 1$\sigma$ region, suggesting a modest but noticeable deviation from the LESC model at earlier epochs.

Fig~\ref{fig_5} shows the evolution of the dimensionless matter density profile $r(z)$. At low redshifts, the LESC model closely follows the $\Lambda$CDM prediction, with both models producing nearly identical results. However, as redshift increases, the LESC prediction deviates noticeably from the standard $(1 + z)^3$ scaling expected in the concordance model, with the $\Lambda$CDM curve moving outside the 1$\sigma$ and eventually 2$\sigma$ confidence regions of the LESC model. This indicates a significant difference in matter density evolution at earlier epochs. It is important to point out that the non-metricity contribution in cosmology is entirely described by a function of time, or equivalently, redshift.

The evolution of this contribution, shown in Fig. \ref{fig_metrcitiy}, remains positive across cosmic history. In the recent Universe (up to \(z = 0.5\)), it is monotonically decreasing, while in the early Universe, it was an increasing function. This contribution could be interpreted as an effective cosmological constant. A major difference between the present approach and the $\Lambda$CDM model is the non-conservation of the energy-momentum tensor. While in $\Lambda$CDM this is guaranteed, in Yano-Schrödinger hyperfluid LESC model, only the effective energy-momentum tensor is conserved. From the point of view of irreversible thermodynamics of open systems, the non-conservation, which occurs due to the presence of non-metricity could be related to particle creation and/or annihilation processes, in a similar way as proposed by Prigogine \cite{prigogine1988thermodynamics,prigogine1989thermodynamics}. Hence, the non-metricity contributions could interact with the ordinary matter part, and effectively behave like interacting dark-energy models. The thermodynamical investigations go beyond the scope of this paper, but could be interesting to study in a follow-up.

To quantitatively assess the performance of the LESC model relative to the standard $\Lambda$CDM model, we compute the logarithm of the Bayes factor, $\ln(B_{ij})$. For our analysis, we find: $\ln(B_{ij}) = 3.8235$ According to Jeffreys’ scale, this value provides moderate evidence in favor of the LESC model over the $\Lambda$CDM model. This result indicates that the LESC model not only fits the data well but also offers a balanced trade off between model fit and complexity.

Additionally, the comparison of the minimum chi-square values, $\chi^2_{\text{min}}$, provides further insight into the goodness of fit. The LESC model yields $\chi^2_{\text{min}} = 1553.79$, which is lower than that of the standard $\Lambda$CDM model, for which $\chi^2_{\text{min}} = 1578.92$. This lower value of $\chi^2_{\text{min}}$ shows that the LESC model fits the combined data better than the $\Lambda$CDM model, supporting its potential as a strong alternative cosmological model.
\section{Conclusion}\label{sec_5}
In this work, we have investigated the FLRW cosmology of the Yano-Schrödinger hyperfluid, a natural extension of the perfect fluid concept used in General Relativity. Unlike standard cosmological models such as $\Lambda$CDM, which focus mainly on spacetime curvature through the Einstein-Hilbert action, the YS hyperfluid incorporates nonmetricity sourced by a specific type of hypermomentum. This nonmetricity introduces new geometric features that affect the dynamics of energy and matter, modifying cosmic expansion and potentially offering new insights into dark energy. These contributions, often neglected in simpler frameworks, reveal underlying mechanisms that govern the Universe’s accelerated expansion. The YS framework provides a self-consistent incorporation of these effects while preserving stability and causality in the field equations, thereby offering a richer structure to explore the link between nonmetricity and dark energy.

From an observational perspective, we propose the LESC model, where effective nonmetricity contributions to pressure and matter density are linearly related as $p_{\text{eff}} = \omega \rho_{\text{eff}}$. We conduct a thorough comparative analysis with the $\Lambda$CDM model using multiple cosmological probes. Our results show that both models fit current observational data well. The LESC model’s constraints on $H_0$ and the sound horizon $r_d$ are broadly consistent with Planck’s measurements, albeit with slightly larger uncertainties due to the full systematic treatment in Cosmic Chronometer data. The evolution of key cosmological quantities such as $H(z)$, distance modulus $\mu(z)$, the deceleration parameter $q(z)$, jerk parameter $j(z)$, and the $Om(z)$ diagnostic reveal subtle but statistically significant deviations from $\Lambda$CDM at higher redshifts, particularly related to nonmetricity effects.

Importantly, our analysis carefully examines the 1$\sigma$ (68\%) and 2$\sigma$  (95\%) confidence intervals for these parameters and diagnostics. We find that at low redshifts, the $\Lambda$CDM predictions generally lie within the 1$\sigma$ confidence region of the LESC model, indicating strong statistical consistency between the two. However, at higher redshifts, some parameters and diagnostics fall outside the 1$\sigma$ but remain mostly within the 2$\sigma$ region, suggesting mild yet notable deviations.

Statistically, the logarithm of the Bayes factor, $\ln B_{ij} \approx 3.82$, indicates moderate evidence in favor of the LESC model over the $\Lambda$CDM framework. Furthermore, the LESC model yields a lower minimum chi-square value, $\chi^2_{\text{min}} = 1553.79$, compared to 1578.92 for $\Lambda$CDM, suggesting a better overall consistency with the observational data.

Our study further reveals that the nonmetricity contribution, expressed as a function of redshift, remains positive throughout cosmic history, decreasing monotonically in the recent Universe and increasing in the early epochs. This behavior can be interpreted as an effective cosmological constant component, distinguishing the YS hyperfluid approach from $\Lambda$CDM, where energy-momentum conservation is exact. In contrast, the LESC model allows for non-conservation of the energy-momentum tensor due to nonmetricity, which could be linked to particle creation or annihilation processes in the spirit of irreversible thermodynamics. This suggests the possibility of effective interactions between dark energy and matter, an avenue for future exploration.

Looking ahead, we plan to implement the LESC model within the CLASS Boltzmann solver and MontePython for a more comprehensive parameter estimation and model comparison. We also intend to study the model’s implications for large-scale structure formation using the ME-Gadget-4 N-body simulation code, which will test its predictions against observations of cosmic growth and clustering.
\begin{acknowledgements}
   S.H. acknowledges the support of National Natural Science Foundation of China under Grants No. W2433018 and No. 11675143, and the National Key Reserach and Development Program of China under Grant No. 2020YFC2201503, and thank to  L.Cs. for fruitful discussions.
\end{acknowledgements}

\bibliographystyle{ieeetr}
\bibliography{mybib,mybib2,mybib_1}

\end{document}